\documentclass[twocolumn,prl,superscriptaddress,showpacs,preprintnumbers,amsmath,amssymb]{revtex4}

\usepackage{graphicx}
\usepackage{dcolumn}
\usepackage{bm}
\usepackage{color}

\begin{document}
\noindent {\bf Ikeda and Miyazaki Reply:} 
In the preceding comment~\cite{Schilling2010}, Schilling and Schmid (SS)
criticized our recent statement 
that reconsideration and revision of mode coupling theory (MCT) from the ground up are 
in order~\cite{Ikeda2010}. 
Our statement is based on the fact that 
long time limit of self-part of van Hove
function $G_{c,\infty}^{(s)}(r;d)$ 
exhibits unphysical negative dips in as low dimension as $d=6$ and the
depths of the dips increase with dimension~\cite{Ikeda2010}.  
Although SS agreed with the existence of the dips, they suggested that
these results are not yet sufficient to verify 
the above statement. 
Their suggestion is based upon two observations: 
(i) The negative dip of $G_{c,\infty}^{(s)}(r;d)$ is demonstrated to
decrease as $d$ increases. 
It is also shown that the dips are very sensitive to the shape of the
nonergodicity parameter $f_c^{(s)}(k;d)$ and that an analytic function
which fits $f_c^{(s)}(k;d)$ reasonably well can eliminate the negative dip. 
They also claim that such a glitch is not surprising given the nonlinear
structure of MCT and may not affect the quality of the theory.
(ii) The long-time and large $d$ limit should be taken with caution. 
Numerical studies on a long-ranged $\phi^4$-model\cite{Kob1991} show that the limits of 
$t\rightarrow \infty$ and $N\rightarrow \infty$ are not commutable,
where $N$ is the number of interaction bonds which should be
proportional to $d$ for off-lattice liquids. 

\begin{figure}[h]
\begin{center}
\includegraphics[width=1.0\columnwidth]{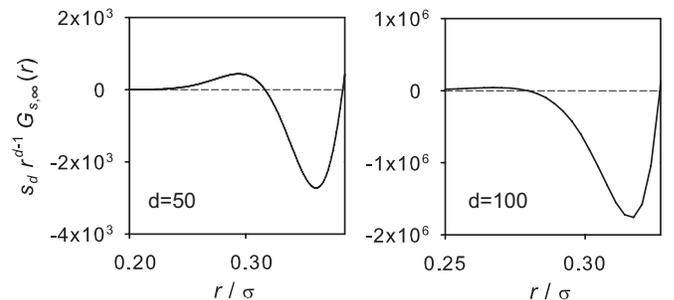}
\caption{
$s_dr^{d-1}G_{s,\infty}(r)$ for $d=50$ (left) and 100 (right), where
$s_d$ is the surface area of the $d$-dimensional unit sphere. 
}
\label{fig_ikeda}
\end{center}
\end{figure}

Regarding the point (i), we believe that our conclusion is not affected by their objection. 
In Figure \ref{fig_ikeda}, we show $G_{c,\infty}^{(s)}(r;d)$ multiplied by
$r^{d-1}$, or the probability density function, 
in much higher dimensions than 
those shown in Ref.~\cite{Ikeda2010}.
Similar dips have been observed for the collective part of the van Hove
function $G_{c,\infty}^{(c)}(r;d)$. 
It clearly demonstrates that the amplitudes of the negative dips increase as $d$ increases.
The negative dips remain noticeable even at $d =100$ where  
the asymptotic scaling of MCT critical point $\varphi_{mct}(d) \propto d^2/2^{d}$ 
is observed~\cite{Ikeda2010,Schmid2010}).   
Note also that  the probability density $r^{d-1}G_{c,\infty}^{(s)}(r;d)$ 
rather than $G_{c,\infty}^{(s)}(r;d)$ is a natural observables, 
because the latter, if negative or positive, becomes negligibly small
for arbitrary non-zero $r$'s as $d$ increases.
We emphasize that both the peculiar $d$-dependence of $\varphi_{mct}(d)$ 
and the pathological negative dips of $G_{c,\infty}^{(s)}(r;d)$ derive
from the non-Gaussian shape of $f_c^{(s)}(k;d)$.  
One can not ignore this non-Gaussianity as a minor glitch hidden
in MCT because the different scaling of $\varphi_{mct}(d)$ from the
prediction of replica theory is a grave problem which may undermine the
mean-field picture of the glass transition theory.

Regarding the second point (ii), 
we believe that our argument is irrelevant to 
the conclusions of Ref.\cite{Kob1991} where  
the relaxation time of a long-ranged model 
was demonstrated to increase with the system size $N$ and diverge in thermodynamic limit
$N\rightarrow \infty$ in the ordered phase. 
This thermodynamic limit should not be taken to be equivalent to
large-$d$ limit in our MCT analysis;
MCT as well as the replica theory are 
``mean-field theories'' {\it a la spin-glasses} even in {\it finite} $d$'s.  
What we have discussed in our paper is the inconsistencies between 
the two theories in large but finite $d$'s (and not 
the effects of the  finite-size nor finite-$d$ in conventional critical phenomena).
The reason to consider large $d$'s was to 
avoid unwanted approximations for the static parameters.

Finally, we agree with SS in that our conclusions would not imply that
MCT is invalid in a certain time and temperature regime for two- and
three-dimension liquids. 
One can not overstate the success of MCT in predicting dynamical
properties of the supercooled liquids. 
The question is why MCT is so powerful and robust, at least in low dimensions, 
despite of 
many uncontrolled and controversial approximations in its derivation. 
Another question is in what sense MCT is the mean-field theory, if 
there really exists a mean-field theory of the glass transition.  
Demonstrating the situations where MCT breaks down and fixing the breakdowns
would help us to answer to these important but unanswered questions.
This is the reason why we claim that 
``reconsideration and revision of MCT from the
ground up are in order''. \\

\noindent Atsushi Ikeda and Kunimasa Miyazaki \\
\indent Institute of Physics, University of Tsukuba, \\
\indent Tennodai 1-1-1, Tsukuba 305-8571, Japan


\begin{thebibliography}{8}
\expandafter\ifx\csname natexlab\endcsname\relax\def\natexlab#1{#1}\fi
\expandafter\ifx\csname bibnamefont\endcsname\relax
  \def\bibnamefont#1{#1}\fi
\expandafter\ifx\csname bibfnamefont\endcsname\relax
  \def\bibfnamefont#1{#1}\fi
\expandafter\ifx\csname citenamefont\endcsname\relax
  \def\citenamefont#1{#1}\fi
\expandafter\ifx\csname url\endcsname\relax
  \def\url#1{\texttt{#1}}\fi
\expandafter\ifx\csname urlprefix\endcsname\relax\def\urlprefix{URL }\fi
\providecommand{\bibinfo}[2]{#2}
\providecommand{\eprint}[2][]{\url{#2}}

\bibitem{Schilling2010}
\bibinfo{author}{\bibfnamefont{R.}~\bibnamefont{Schilling}} \bibnamefont{and}
\bibinfo{author}{\bibfnamefont{B.}~\bibnamefont{Schmid}}, preceding comment. 

\bibitem{Ikeda2010}
\bibinfo{author}{\bibfnamefont{A.}~\bibnamefont{Ikeda}} \bibnamefont{and}
\bibinfo{author}{\bibfnamefont{K.}~\bibnamefont{Miyazaki}}, 
\bibinfo{journal}{Phys. Rev. Lett.} \textbf{\bibinfo{volume}{{\bf 104}}},
\bibinfo{pages}{255704} (\bibinfo{year}{2010}).

\bibitem{Kob1991}
\bibinfo{author}{\bibfnamefont{W.}~\bibnamefont{Kob}} \bibnamefont{and}
\bibinfo{author}{\bibfnamefont{R.}~\bibnamefont{Schilling}},
\bibinfo{journal}{J. Phys.: Condens. Matter} \textbf{\bibinfo{volume}{{\bf 3}}},
\bibinfo{pages}{9195} (\bibinfo{year}{1991}).

\bibitem{Schmid2010}
\bibinfo{author}{\bibfnamefont{B.}~\bibnamefont{Schmid}} \bibnamefont{and}
\bibinfo{author}{\bibfnamefont{R.}~\bibnamefont{Schilling}},
\bibinfo{journal}{Phys. Rev. {\rm E}} \textbf{\bibinfo{volume}{{\bf 81}}},
\bibinfo{pages}{041502} (\bibinfo{year}{2010}).

\end{thebibliography}
\end{document}